\newcommand{\pkg}[1]{\texttt{#1}}
\newcommand{\proglang}[1]{\texttt{#1}}

\newcommand{\JSON}{\texttt{JSON}\xspace}
\newcommand{\toJSON}{\texttt{toJSON}\xspace}
\newcommand{\fromJSON}{\texttt{fromJSON}\xspace}
\newcommand{\XML}{\texttt{XML}\xspace}
\newcommand{\jsonlite}{\texttt{jsonlite}\xspace}

\documentclass{article}\usepackage[]{graphicx}\usepackage[]{color}
\makeatletter
\def\maxwidth{ %
  \ifdim\Gin@nat@width>\linewidth
    \linewidth
  \else
    \Gin@nat@width
  \fi
}
\makeatother

\definecolor{fgcolor}{rgb}{0.345, 0.345, 0.345}
\newcommand{\hlnum}[1]{\textcolor[rgb]{0.686,0.059,0.569}{#1}}%
\newcommand{\hlstr}[1]{\textcolor[rgb]{0.192,0.494,0.8}{#1}}%
\newcommand{\hlcom}[1]{\textcolor[rgb]{0.678,0.584,0.686}{\textit{#1}}}%
\newcommand{\hlopt}[1]{\textcolor[rgb]{0,0,0}{#1}}%
\newcommand{\hlstd}[1]{\textcolor[rgb]{0.345,0.345,0.345}{#1}}%
\newcommand{\hlkwb}[1]{\textcolor[rgb]{0.69,0.353,0.396}{#1}}%
\newcommand{\hlkwc}[1]{\textcolor[rgb]{0.333,0.667,0.333}{#1}}%
\newcommand{\hlkwd}[1]{\textcolor[rgb]{0.737,0.353,0.396}{\textbf{#1}}}%

\usepackage{framed}
\makeatletter
\newenvironment{kframe}{%
 \def\at@end@of@kframe{}%
 \ifinner\ifhmode%
  \def\at@end@of@kframe{\end{minipage}}%
  \begin{minipage}{\columnwidth}%
 \fi\fi%
 \def\FrameCommand##1{\hskip\@totalleftmargin \hskip-\fboxsep
 \colorbox{shadecolor}{##1}\hskip-\fboxsep
     \hskip-\linewidth \hskip-\@totalleftmargin \hskip\columnwidth}%
 \MakeFramed {\advance\hsize-\width
   \@totalleftmargin\z@ \linewidth\hsize
   \@setminipage}}%
 {\par\unskip\endMakeFramed%
 \at@end@of@kframe}
\makeatother

\definecolor{shadecolor}{rgb}{.97, .97, .97}
\definecolor{messagecolor}{rgb}{0, 0, 0}
\definecolor{warningcolor}{rgb}{1, 0, 1}
\definecolor{errorcolor}{rgb}{1, 0, 0}
\newenvironment{knitrout}{}{} 

\usepackage{alltt}
\author{Jeroen Ooms\\UCLA Department of Statistics} 
\date{}
\title{The \jsonlite Package: A Practical and Consistent Mapping Between JSON Data and R Objects}

\usepackage{url}
\usepackage{fullpage}
\usepackage{xspace}
\usepackage{booktabs}
\usepackage{enumitem}
\usepackage[hidelinks]{hyperref}
\usepackage[round]{natbib}
\usepackage{fancyvrb}
\usepackage[toc,page]{appendix}
\usepackage{breakurl}

\usepackage{parskip}

\usepackage{setspace}
\IfFileExists{upquote.sty}{\usepackage{upquote}}{}
\begin{document}

\maketitle

\begin{abstract}
A naive realization of \JSON data in R maps \JSON \emph{arrays} to an \texttt{unnamed list}, and \JSON \emph{objects} to a \texttt{named list}. However, in practice a list is an awkward, inefficient type to store and manipulate data. Most statistical applications work with (homogeneous) vectors, matrices or data frames. Therefore \JSON packages in R typically define certain special cases of \JSON structures which map to simpler R types. Currently there exist no formal guidelines, or even consensus between implementations on how R data should be represented in \JSON. Furthermore, upon closer inspection, even the most basic data structures in R actually do not perfectly map to their \JSON counterparts and leave some ambiguity for edge cases. These problems have resulted in different behavior between implementations and can lead to unexpected output. This paper explicitly describes a mapping between R classes and \JSON data, highlights potential problems, and proposes conventions that generalize the mapping to cover all common structures. We emphasize the importance of type consistency when using \JSON to exchange dynamic data, and illustrate using examples and anecdotes. The \jsonlite R package is used throughout the paper as a reference implementation.
\end{abstract}

\section{Introduction}

\emph{JavaScript Object Notation} (\JSON) is a text format for the serialization of structured data \citep{crockford2006application}. It is derived from the object literals of \proglang{JavaScript}, as defined in the \proglang{ECMAScript} Programming Language Standard, Third Edition \citep{ecma1999262}. Design of \JSON is simple and concise in comparison with other text based formats, and it was originally proposed by Douglas Crockford as a ``fat-free alternative to \XML'' \citep{crockford2006json}. The syntax is easy for humans to read and write, easy for machines to parse and generate and completely described in a single page at \url{http://www.json.org}. The character encoding of \JSON text is always Unicode, using \texttt{UTF-8} by default \citep{crockford2006application}, making it naturally compatible with non-latin alphabets. Over the past years, \JSON has become hugely popular on the internet as a general purpose data interchange format. High quality parsing libraries are available for almost any programming language, making it easy to implement systems and applications that exchange data over the network using JSON. For R \citep{R}, several packages that assist the user in generating, parsing and validating JSON are available through CRAN, including \pkg{rjson} \citep{rjson}, \pkg{RJSONIO} \citep{RJSONIO}, and \pkg{jsonlite} \citep{jsonlite}.

The emphasis of this paper is not on discussing the \JSON format or any particular implementation for using \JSON with R. We refer to \cite{nolan2014xml} for a comprehensive introduction, or one of the many tutorials available on the web. Instead we take a high level view and discuss how R data structures are most naturally represented in \JSON. This is not a trivial problem, particulary for complex or relational data as they frequently appear in statistical applications. Several R packages implement \toJSON and \fromJSON functions which directly convert R objects into \JSON and vice versa. However, the exact mapping between the various R data classes \JSON structures is not self evident. Currently, there are no formal guidelines, or even consensus between implementations on how R data should be represented in \JSON. Furthermore, upon closer inspection, even the most basic data structures in R actually do not perfectly map to their \JSON counterparts, and leave some ambiguity for edge cases. These problems have resulted in different behavior between implementations, and can lead to unexpected output for certain special cases. To further complicate things, best practices of representing data in \JSON have been established outside the R community. Incorporating these conventions where possible is important to maximize interoperability.


\subsection{Parsing and type safety}

The \texttt{JSON} format specifies 4 primitive types (\texttt{string}, \texttt{number}, \texttt{boolean}, \texttt{null}) and two \emph{universal structures}:

\begin{itemize} 
  \item A \JSON \emph{object}: an unordered collection of zero or more name/value
   pairs, where a name is a string and a value is a string, number,
   boolean, null, object, or array.
  \item A \JSON \emph{array}: an ordered sequence of zero or more values.
\end{itemize}

\noindent Both these structures are heterogeneous; i.e. they are allowed to contain elements of different types. Therefore, the native R realization of these structures is a \texttt{named list} for \JSON objects, and \texttt{unnamed list} for \JSON arrays. However, in practice a list is an awkward, inefficient type to store and manipulate data in R. Most statistical applications work with (homogeneous) vectors, matrices or data frames. In order to give these data structures a \JSON representation, we can define certain special cases of \JSON structures which get parsed into other, more specific R types. For example, one convention which all current implementations have in common is that a homogeneous array of primitives gets parsed into an \texttt{atomic vector} instead of a \texttt{list}. The \pkg{RJSONIO} documentation uses the term ``simplify'' for this, and we adopt this jargon.

\begin{knitrout}
\definecolor{shadecolor}{rgb}{0.969, 0.969, 0.969}\color{fgcolor}\begin{kframe}
\begin{alltt}
\hlstd{txt} \hlkwb{<-} \hlstr{"[12, 3, 7]"}
\hlstd{x} \hlkwb{<-} \hlkwd{fromJSON}\hlstd{(txt)}
\hlkwd{is}\hlstd{(x)}
\end{alltt}
\begin{verbatim}
[1] "numeric" "vector" 
\end{verbatim}
\begin{alltt}
\hlkwd{print}\hlstd{(x)}
\end{alltt}
\begin{verbatim}
[1] 12  3  7
\end{verbatim}
\end{kframe}
\end{knitrout}

This seems very reasonable and it is the only practical solution to represent vectors in \JSON. However the price we pay is that automatic simplification can compromise type-safety in the context of dynamic data. For example, suppose an R package uses \fromJSON to pull data from a JSON API on the web, similar to the example above. However, for some particular combination of parameters, the result includes a \texttt{null} value, e.g: \texttt{[12, null, 7]}. This is actually quite common, many APIs use \texttt{null} for missing values or unset fields. This case makes the behavior of parsers ambiguous, because the \JSON array is technically no longer homogenous. And indeed, some implementations will now return a \texttt{list} instead of a \texttt{vector}. If the user had not anticipated this scenario and the script assumes a \texttt{vector}, the code is likely to run into type errors. 

The lesson here is that we need to be very specific and explicit about the mapping that is implemented to convert between \JSON and R objects. When relying on \JSON as a data interchange format, the behavior of the parser must be consistent and unambiguous. Clients relying on \JSON to get data in and out of R must know exactly what to expect in order to facilitate reliable communication, even if the data themselves are dynamic. Similarly, R code using dynamic \JSON data from an external source is only reliable when the conversion from \JSON to R is consistent. Moreover a practical mapping must incorporate existing conventions and uses the most natural representation of certain structures in R. For example, we could argue that instead of falling back on a \texttt{list}, the array above is more naturally interpreted as a numeric vector where the \texttt{null} becomes a missing value (\texttt{NA}). These principles will extrapolate as we start discussing more complex \JSON structures representing matrices and data frames.

%

\subsection{Reference implementation: the \jsonlite package}

The \jsonlite package provides a reference implementation of the conventions proposed in this document. \jsonlite is a fork of the \texttt{RJSONIO} package by Duncan Temple Lang, which again builds on \texttt{libjson} \texttt{C++} library from Jonathan Wallace. The \jsonlite package uses the parser from \texttt{RJSONIO}, but the R code has been rewritten from scratch. Both packages implement \toJSON and \fromJSON functions, but their output is quite different. Finally, the \jsonlite package contains a large set of unit tests to validate that R objects are correctly converted to \JSON and vice versa. These unit tests cover all classes and edge cases mentioned in this document, and could be used to validate if other implementations follow the same conventions. 

\begin{knitrout}
\definecolor{shadecolor}{rgb}{0.969, 0.969, 0.969}\color{fgcolor}\begin{kframe}
\begin{alltt}
\hlkwd{library}\hlstd{(testthat)}
\hlkwd{test_package}\hlstd{(}\hlstr{"jsonlite"}\hlstd{)}
\end{alltt}
\end{kframe}
\end{knitrout}

Note that even though \texttt{JSON} allows for inserting arbitrary white space and indentation, the unit tests assume that white space is trimmed.

\subsection{Class-based versus type-based encoding}
\label{serializejson}

The \jsonlite package actually implements two systems for translating between R objects and \JSON. This document focuses on the \toJSON and \fromJSON functions which use R's class-based method dispatch. For all of the common classes in R, the \jsonlite package implements \toJSON methods as described in this document. Users in R can extend this system by implementing additional methods for other classes. However this also means that classes that do not have the \toJSON method defined are not supported. Furthermore, the implementation of a specific \toJSON method determines which data and metadata in the objects of this class gets encoded in its \JSON representation, and how. In this respect, \toJSON is similar to e.g. the \texttt{print} function, which also provides a certain \emph{representation} of an object based on its class and optionally some print parameters. This representation does not necessarily reflect all information stored in the object, and there is no guaranteed one-to-one correspondence between R objects and \JSON. I.e. calling \texttt{fromJSON(toJSON(object))} will return an object which only contains the data that was encoded by the \toJSON method for this particular class, and which might even have a different class than the original.

The alternative to class-based method dispatch is to use type-based encoding, which \jsonlite implements in the functions \texttt{serializeJSON} and \texttt{unserializeJSON}. All data structures in R get stored in memory using one of the internal \texttt{SEXP} storage types, and \texttt{serializeJSON} defines an encoding schema which captures the type, value, and attributes for each storage type. The result is \JSON output which closely resembles the internal structure of the underlying C data types, and which can be perfectly restored to the original R object using \texttt{unserializeJSON}. This system is relatively straightforward to implement, however the disadvantage is that the resulting \JSON is very verbose, hard to interpret, and cumbersome to generate in the context of another language or system. For most applications this is actually impractical because it requires the client/consumer to understand and manipulate R data types, which is difficult and reduces interoperability. Instead we can make data in R more accessible to third parties by defining sensible \JSON representations that are natural for the class of an object, rather than its internal storage type. This document does not discuss the \texttt{serializeJSON} system in any further detail, and solely treats the class based system implemented in \toJSON and \fromJSON. However the reader that is interested in full serialization of R objects into \JSON is encouraged to have a look at the respective manual pages.

\subsection{Scope and limitations}

Before continuing, we want to stress some limitations of encoding R data structures in \JSON. Most importantly, there are the limitations to types of objects that can be represented. In general, temporary in-memory properties such as connections, file descriptors and (recursive) memory references are always difficult if not impossible to store in a sensible way, regardless of the language or serialization method. This document focuses on the common R classes that hold \emph{data}, such as vectors, factors, lists, matrices and data frames. We do not treat language level constructs such as expressions, functions, promises, which hold little meaning outside the context of R. We also don't treat special compound classes such as linear models or custom classes defined in contributed packages. When designing systems or protocols that interact with R, it is highly recommended to stick with the standard data structures for the interface input/output.

Then there are limitations introduced by the format. Because \JSON is a human readable, text-based format, it does not support binary data, and numbers are stored in their decimal notation. The latter leads to loss of precision for real numbers, depending on how many digits the user decides to print. Several dialects of \JSON exists such as \texttt{BSON} \citep{chodorow2013mongodb} or \texttt{MSGPACK} \citep{msgpack}, which extend the format with various binary types. However, these formats are much less popular, less interoperable, and often impractical, precisely because they require binary parsing and abandon human readability. The simplicity of \JSON is what makes it an accessible and widely applicable data interchange format. In cases where it is really needed to include some binary data in \JSON, one can use something like \texttt{base64} to encode it as a string.

Finally, as mentioned earlier, \fromJSON is not a perfect inverse function of \toJSON, as would be the case for \texttt{serialializeJSON} and \texttt{unserializeJSON}. The class based mappings are designed for concise and practical encoding of the various common data structures. Our implementation of \toJSON and \fromJSON approximates a reversible mapping between R objects and \JSON for the standard data classes, but there are always limitations and edge cases. For example, the \JSON representation of an empty vector, empty list or empty data frame are all the same: \texttt{"[ ]"}. Also some special vector types such as factors, dates or timestamps get coerced to strings, as they would in for example \texttt{CSV}. This is a quite typical and expected behavior among text based formats, but it does require some additional interpretation on the consumer side.

\section{Converting between \JSON and R classes}
 
This section lists examples of how the common R classes are represented in \JSON. As explained before, the \toJSON function relies on method dispatch, which means that objects get encoded according to their \texttt{class}. If an object has multiple \texttt{class} values, R uses the first occurring class which has a \toJSON method. If none of the classes of an object has a \toJSON method, an error is raised.

\subsection{Atomic vectors}

The most basic data type in R is the atomic vector. The atomic vector holds an ordered set of homogeneous values of type \texttt{"logical"} (booleans), \texttt{character} (strings), \texttt{"raw"} (bytes), \texttt{numeric} (doubles), \texttt{"complex"} (complex numbers with a real and imaginary part), or \texttt{integer}. Because R is fully vectorized, there is no user level notion of a primitive: a scalar value is considered a vector of length 1. Atomic vectors map to \JSON arrays:

\begin{knitrout}
\definecolor{shadecolor}{rgb}{0.969, 0.969, 0.969}\color{fgcolor}\begin{kframe}
\begin{alltt}
\hlstd{x} \hlkwb{<-} \hlkwd{c}\hlstd{(}\hlnum{1}\hlstd{,} \hlnum{2}\hlstd{, pi)}
\hlkwd{cat}\hlstd{(}\hlkwd{toJSON}\hlstd{(x))}
\end{alltt}
\begin{verbatim}
[ 1, 2, 3.14 ]
\end{verbatim}
\end{kframe}
\end{knitrout}

The \JSON array is the only appropriate structure to encode a vector, however note that vectors in R are homogeneous, whereas the \JSON array is actually heterogeneous, but \JSON does not make this distinction.

\subsubsection{Missing values}

A typical domain specific problem when working with statistical data is presented by missing values: a concept foreign to many other languages. Besides regular values, each vector type in R except for \texttt{raw} can hold \texttt{NA} as a value. Vectors of type \texttt{double} and \texttt{complex} define three additional types of non finite values: \texttt{NaN}, \texttt{Inf} and \texttt{-Inf}. The \JSON format does not natively support any of these types; therefore such values values need to be encoded in some other way. There are two obvious approaches. The first one is to use the  \JSON \texttt{null} type. For example:

\begin{knitrout}
\definecolor{shadecolor}{rgb}{0.969, 0.969, 0.969}\color{fgcolor}\begin{kframe}
\begin{alltt}
\hlstd{x} \hlkwb{<-} \hlkwd{c}\hlstd{(}\hlnum{TRUE}\hlstd{,} \hlnum{FALSE}\hlstd{,} \hlnum{NA}\hlstd{)}
\hlkwd{cat}\hlstd{(}\hlkwd{toJSON}\hlstd{(x))}
\end{alltt}
\begin{verbatim}
[ true, false, null ]
\end{verbatim}
\end{kframe}
\end{knitrout}

The other option is to encode missing values as strings by wrapping them in double quotes:

\begin{knitrout}
\definecolor{shadecolor}{rgb}{0.969, 0.969, 0.969}\color{fgcolor}\begin{kframe}
\begin{alltt}
\hlstd{x} \hlkwb{<-} \hlkwd{c}\hlstd{(}\hlnum{1}\hlstd{,} \hlnum{2}\hlstd{,} \hlnum{NA}\hlstd{,} \hlnum{NaN}\hlstd{,} \hlnum{Inf}\hlstd{,} \hlnum{10}\hlstd{)}
\hlkwd{cat}\hlstd{(}\hlkwd{toJSON}\hlstd{(x))}
\end{alltt}
\begin{verbatim}
[ 1, 2, "NA", "NaN", "Inf", 10 ]
\end{verbatim}
\end{kframe}
\end{knitrout}

Both methods result in valid \JSON, but both have a limitation: the problem with the \texttt{null} type is that it is impossible to distinguish between different types of missing data, which could be a problem for numeric vectors. The values \texttt{Inf}, \texttt{-Inf}, \texttt{NA} and \texttt{NaN} carry different meanings, and these should not get lost in the encoding. However, the problem with encoding missing values as strings is that this method can not be used for character vectors, because the consumer won't be able to distinguish the actual string \texttt{"NA"} and the missing value \texttt{NA}. This would create a likely source of bugs, where clients mistakenly interpret \texttt{"NA"} as an actual value, which is a common problem with text-based formats such as CSV. For this reason, \jsonlite uses the following defaults:

\begin{itemize}
 \item Missing values in non-numeric vectors (\texttt{logical}, \texttt{character}) are encoded as \texttt{null}.
 \item Missing values in numeric vectors (\texttt{double}, \texttt{integer}, \texttt{complex}) are encoded as strings.
\end{itemize}

We expect that these conventions are most likely to result in the correct interpretation of missing values. Some examples:

\begin{knitrout}
\definecolor{shadecolor}{rgb}{0.969, 0.969, 0.969}\color{fgcolor}\begin{kframe}
\begin{alltt}
\hlkwd{cat}\hlstd{(}\hlkwd{toJSON}\hlstd{(}\hlkwd{c}\hlstd{(}\hlnum{TRUE}\hlstd{,} \hlnum{NA}\hlstd{,} \hlnum{NA}\hlstd{,} \hlnum{FALSE}\hlstd{)))}
\end{alltt}
\begin{verbatim}
[ true, null, null, false ]
\end{verbatim}
\begin{alltt}
\hlkwd{cat}\hlstd{(}\hlkwd{toJSON}\hlstd{(}\hlkwd{c}\hlstd{(}\hlstr{"FOO"}\hlstd{,} \hlstr{"BAR"}\hlstd{,} \hlnum{NA}\hlstd{,} \hlstr{"NA"}\hlstd{)))}
\end{alltt}
\begin{verbatim}
[ "FOO", "BAR", null, "NA" ]
\end{verbatim}
\begin{alltt}
\hlkwd{cat}\hlstd{(}\hlkwd{toJSON}\hlstd{(}\hlkwd{c}\hlstd{(}\hlnum{3.14}\hlstd{,} \hlnum{NA}\hlstd{,} \hlnum{NaN}\hlstd{,} \hlnum{21}\hlstd{,} \hlnum{Inf}\hlstd{,} \hlopt{-}\hlnum{Inf}\hlstd{)))}
\end{alltt}
\begin{verbatim}
[ 3.14, "NA", "NaN", 21, "Inf", "-Inf" ]
\end{verbatim}
\begin{alltt}
\hlkwd{cat}\hlstd{(}\hlkwd{toJSON}\hlstd{(}\hlkwd{c}\hlstd{(}\hlnum{3.14}\hlstd{,} \hlnum{NA}\hlstd{,} \hlnum{NaN}\hlstd{,} \hlnum{21}\hlstd{,} \hlnum{Inf}\hlstd{,} \hlopt{-}\hlnum{Inf}\hlstd{),} \hlkwc{na} \hlstd{=} \hlstr{"null"}\hlstd{))}
\end{alltt}
\begin{verbatim}
[ 3.14, null, null, 21, null, null ]
\end{verbatim}
\end{kframe}
\end{knitrout}

\subsubsection{Special vector types: dates, times, factor, complex}

Besides missing values, \JSON also lacks native support for some of the basic vector types in R that frequently appear in data sets. These include vectors of class \texttt{Date}, \texttt{POSIXt} (timestamps), \texttt{factors} and \texttt{complex} vectors. By default, the \jsonlite package coerces these types to strings (using \texttt{as.character}):

\begin{knitrout}
\definecolor{shadecolor}{rgb}{0.969, 0.969, 0.969}\color{fgcolor}\begin{kframe}
\begin{alltt}
\hlkwd{cat}\hlstd{(}\hlkwd{toJSON}\hlstd{(}\hlkwd{Sys.time}\hlstd{()} \hlopt{+} \hlnum{1}\hlopt{:}\hlnum{3}\hlstd{))}
\end{alltt}
\begin{verbatim}
[ "2014-03-11 21:16:05", "2014-03-11 21:16:06", "2014-03-11 21:16:07" ]
\end{verbatim}
\begin{alltt}
\hlkwd{cat}\hlstd{(}\hlkwd{toJSON}\hlstd{(}\hlkwd{as.Date}\hlstd{(}\hlkwd{Sys.time}\hlstd{())} \hlopt{+} \hlnum{1}\hlopt{:}\hlnum{3}\hlstd{))}
\end{alltt}
\begin{verbatim}
[ "2014-03-13", "2014-03-14", "2014-03-15" ]
\end{verbatim}
\begin{alltt}
\hlkwd{cat}\hlstd{(}\hlkwd{toJSON}\hlstd{(}\hlkwd{factor}\hlstd{(}\hlkwd{c}\hlstd{(}\hlstr{"foo"}\hlstd{,} \hlstr{"bar"}\hlstd{,} \hlstr{"foo"}\hlstd{))))}
\end{alltt}
\begin{verbatim}
[ "foo", "bar", "foo" ]
\end{verbatim}
\begin{alltt}
\hlkwd{cat}\hlstd{(}\hlkwd{toJSON}\hlstd{(}\hlkwd{complex}\hlstd{(}\hlkwc{real} \hlstd{=} \hlkwd{runif}\hlstd{(}\hlnum{3}\hlstd{),} \hlkwc{imaginary} \hlstd{=} \hlkwd{rnorm}\hlstd{(}\hlnum{3}\hlstd{))))}
\end{alltt}
\begin{verbatim}
[ "0.5+1.7i", "0-2i", "0.37-0.13i" ]
\end{verbatim}
\end{kframe}
\end{knitrout}

When parsing such \JSON strings, these values will appear as character vectors. In order to obtain the original types, the user needs to manually coerce them back to the desired type using the corresponding \texttt{as} function, e.g. \texttt{as.POSIXct}, \texttt{as.Date}, \texttt{as.factor} or \texttt{as.complex}. In this respect, \JSON is subject to the same limitations as text based formats such as CSV.

\subsubsection{Special cases: vectors of length 0 or 1}

Two edge cases deserve special attention: vectors of length 0 and vectors of length 1. In \jsonlite these are encoded respectively as an empty array, and an array of length 1:

\begin{knitrout}
\definecolor{shadecolor}{rgb}{0.969, 0.969, 0.969}\color{fgcolor}\begin{kframe}
\begin{alltt}
\hlcom{# vectors of length 0 and 1}
\hlkwd{cat}\hlstd{(}\hlkwd{toJSON}\hlstd{(}\hlkwd{vector}\hlstd{()))}
\end{alltt}
\begin{verbatim}
[  ]
\end{verbatim}
\begin{alltt}
\hlkwd{cat}\hlstd{(}\hlkwd{toJSON}\hlstd{(pi))}
\end{alltt}
\begin{verbatim}
[ 3.14 ]
\end{verbatim}
\begin{alltt}
\hlcom{# vectors of length 0 and 1 in a named list}
\hlkwd{cat}\hlstd{(}\hlkwd{toJSON}\hlstd{(}\hlkwd{list}\hlstd{(}\hlkwc{foo} \hlstd{=} \hlkwd{vector}\hlstd{())))}
\end{alltt}
\begin{verbatim}
{ "foo" : [  ] }
\end{verbatim}
\begin{alltt}
\hlkwd{cat}\hlstd{(}\hlkwd{toJSON}\hlstd{(}\hlkwd{list}\hlstd{(}\hlkwc{foo} \hlstd{= pi)))}
\end{alltt}
\begin{verbatim}
{ "foo" : [ 3.14 ] }
\end{verbatim}
\begin{alltt}
\hlcom{# vectors of length 0 and 1 in an unnamed list}
\hlkwd{cat}\hlstd{(}\hlkwd{toJSON}\hlstd{(}\hlkwd{list}\hlstd{(}\hlkwd{vector}\hlstd{())))}
\end{alltt}
\begin{verbatim}
[ [  ] ]
\end{verbatim}
\begin{alltt}
\hlkwd{cat}\hlstd{(}\hlkwd{toJSON}\hlstd{(}\hlkwd{list}\hlstd{(pi)))}
\end{alltt}
\begin{verbatim}
[ [ 3.14 ] ]
\end{verbatim}
\end{kframe}
\end{knitrout}

This might seem obvious but these cases result in very different behavior between different \JSON packages. This is probably caused by the fact that R does not have a scalar type, and some package authors decided to treat vectors of length 1 as if they were a scalar. For example, in the current implementations, both \texttt{RJSONIO} and \texttt{rjson} encode a vector of length one as a \JSON primitive when it appears within a list: 

\begin{knitrout}
\definecolor{shadecolor}{rgb}{0.969, 0.969, 0.969}\color{fgcolor}\begin{kframe}
\begin{alltt}
\hlcom{# Other packages make different choices:}
\hlkwd{cat}\hlstd{(rjson::}\hlkwd{toJSON}\hlstd{(}\hlkwd{list}\hlstd{(}\hlkwc{n} \hlstd{=} \hlkwd{c}\hlstd{(}\hlnum{1}\hlstd{))))}
\end{alltt}
\begin{verbatim}
 {"n":1}
\end{verbatim}
\begin{alltt}
\hlkwd{cat}\hlstd{(rjson::}\hlkwd{toJSON}\hlstd{(}\hlkwd{list}\hlstd{(}\hlkwc{n} \hlstd{=} \hlkwd{c}\hlstd{(}\hlnum{1}\hlstd{,} \hlnum{2}\hlstd{))))}
\end{alltt}
\begin{verbatim}
 {"n":[1,2]}
\end{verbatim}
\end{kframe}
\end{knitrout}

When encoding a single dataset this seems harmless, but in the context of dynamic data this inconsistency is almost guaranteed to cause bugs. For example, imagine an R web service which lets the user fit a linear model and sends back the fitted parameter estimates as a \JSON array. The client code then parses the \JSON, and iterates over the array of coefficients to display them in a GUI. All goes well, until the user decides to fit a model with only one predictor. If the \JSON encoder suddenly returns a primitive value where the client is assuming an array, the application will likely break. Any consumer or client would need to be aware of the special case where the vector becomes a primitive, and explicitly take this exception into account when processing the result. When the client fails to do so and proceeds as usual, it will probably call an iterator or loop method on a primitive value, resulting in the obvious errors. For this reason \jsonlite uses consistent encoding schemes which do not depend on variable object properties such as its length. Hence, a vector is always encoded as an array, even when it is of length 0 or 1.

\subsection{Matrices}

Arguably one of the strongest sides of R is its ability to interface libraries for basic linear algebra subprograms \citep{lawson1979basic} such as LAPACK \citep{anderson1999lapack}. These libraries provide well tuned, high performance implementations of important linear algebra operations to calculate anything from inner products and eigen values to singular value decompositions. These are in turn the building blocks of statistical methods such as linear regression or principal component analysis. Linear algebra methods operate on \emph{matrices}, making the matrix one of the most central data classes in R. Conceptually, a matrix consists of a 2 dimensional structure of homogeneous values. It is indexed using 2 numbers (or vectors), representing the row and column number of the matrix respectively. 

\begin{knitrout}
\definecolor{shadecolor}{rgb}{0.969, 0.969, 0.969}\color{fgcolor}\begin{kframe}
\begin{alltt}
\hlstd{x} \hlkwb{<-} \hlkwd{matrix}\hlstd{(}\hlnum{1}\hlopt{:}\hlnum{12}\hlstd{,} \hlkwc{nrow} \hlstd{=} \hlnum{3}\hlstd{,} \hlkwc{ncol} \hlstd{=} \hlnum{4}\hlstd{)}
\hlkwd{print}\hlstd{(x)}
\end{alltt}
\begin{verbatim}
     [,1] [,2] [,3] [,4]
[1,]    1    4    7   10
[2,]    2    5    8   11
[3,]    3    6    9   12
\end{verbatim}
\begin{alltt}
\hlkwd{print}\hlstd{(x[}\hlnum{2}\hlstd{,} \hlnum{4}\hlstd{])}
\end{alltt}
\begin{verbatim}
[1] 11
\end{verbatim}
\end{kframe}
\end{knitrout}

 A matrix is stored in memory as a single atomic vector with an attribute called \texttt{"dim"} defining the dimensions of the matrix. The product of the dimensions is equal to the length of the vector. 

\begin{knitrout}
\definecolor{shadecolor}{rgb}{0.969, 0.969, 0.969}\color{fgcolor}\begin{kframe}
\begin{alltt}
\hlkwd{attributes}\hlstd{(volcano)}
\end{alltt}
\begin{verbatim}
$dim
[1] 87 61
\end{verbatim}
\begin{alltt}
\hlkwd{length}\hlstd{(volcano)}
\end{alltt}
\begin{verbatim}
[1] 5307
\end{verbatim}
\end{kframe}
\end{knitrout}

 Even though the matrix is stored as a single vector, the way it is printed and indexed makes it conceptually a 2 dimensional structure. In \jsonlite a matrix maps to an array of equal-length subarrays:

\begin{knitrout}
\definecolor{shadecolor}{rgb}{0.969, 0.969, 0.969}\color{fgcolor}\begin{kframe}
\begin{alltt}
\hlstd{x} \hlkwb{<-} \hlkwd{matrix}\hlstd{(}\hlnum{1}\hlopt{:}\hlnum{12}\hlstd{,} \hlkwc{nrow} \hlstd{=} \hlnum{3}\hlstd{,} \hlkwc{ncol} \hlstd{=} \hlnum{4}\hlstd{)}
\hlkwd{cat}\hlstd{(}\hlkwd{toJSON}\hlstd{(x))}
\end{alltt}
\begin{verbatim}
[ [ 1, 4, 7, 10 ], [ 2, 5, 8, 11 ], [ 3, 6, 9, 12 ] ]
\end{verbatim}
\end{kframe}
\end{knitrout}

We expect this representation will be the most intuitive to interpret, also within languages that do not have a native notion of a matrix. Note that even though R stores matrices in \emph{column major} order, \jsonlite encodes matrices in \emph{row major} order. This is a more conventional and intuitive way to represent matrices and is consistent with the row-based encoding of data frames discussed in the next section. When the \JSON string is properly indented (recall that white space and line breaks are optional in \JSON), it looks very similar to the way R prints matrices:

\begin{verbatim}
[ [ 1, 4, 7, 10 ], 
  [ 2, 5, 8, 11 ], 
  [ 3, 6, 9, 12 ] ]
\end{verbatim}

 Because the matrix is implemented in R as an atomic vector, it automatically inherits the conventions mentioned earlier with respect to edge cases and missing values:

\begin{knitrout}
\definecolor{shadecolor}{rgb}{0.969, 0.969, 0.969}\color{fgcolor}\begin{kframe}
\begin{alltt}
\hlstd{x} \hlkwb{<-} \hlkwd{matrix}\hlstd{(}\hlkwd{c}\hlstd{(}\hlnum{1}\hlstd{,} \hlnum{2}\hlstd{,} \hlnum{4}\hlstd{,} \hlnum{NA}\hlstd{),} \hlkwc{nrow} \hlstd{=} \hlnum{2}\hlstd{)}
\hlkwd{cat}\hlstd{(}\hlkwd{toJSON}\hlstd{(x))}
\end{alltt}
\begin{verbatim}
[ [ 1, 4 ], [ 2, "NA" ] ]
\end{verbatim}
\begin{alltt}
\hlkwd{cat}\hlstd{(}\hlkwd{toJSON}\hlstd{(x,} \hlkwc{na} \hlstd{=} \hlstr{"null"}\hlstd{))}
\end{alltt}
\begin{verbatim}
[ [ 1, 4 ], [ 2, null ] ]
\end{verbatim}
\begin{alltt}
\hlkwd{cat}\hlstd{(}\hlkwd{toJSON}\hlstd{(}\hlkwd{matrix}\hlstd{(pi)))}
\end{alltt}
\begin{verbatim}
[ [ 3.14 ] ]
\end{verbatim}
\end{kframe}
\end{knitrout}

\subsubsection{Matrix row and column names}

Besides the \texttt{"dim"} attribute, the matrix class has an additional, optional attribute: \texttt{"dimnames"}. This attribute holds names for the rows and columns in the matrix. However, we decided not to include this information in the default \JSON mapping for matrices for several reasons. First of all, because this attribute is optional, often either row or column names or both are \texttt{NULL}. This makes it difficult to define a practical encoding that covers all cases with and without row and/or column names. Secondly, the names in matrices are mostly there for annotation only; they are not actually used in calculations. The linear algebra subroutines mentioned before completely ignore them, and never include any names in their output. So there is often little purpose of setting names in the first place, other than annotation.

When row or column names of a matrix seem to contain vital information, we might want to transform the data into a more appropriate structure. \cite{tidydata} calls this \emph{``tidying''} the data and outlines best practices on storing statistical data in its most appropriate form. He lists the issue where \emph{``column headers are values, not variable names''} as the most common source of untidy data. This often happens when the structure is optimized for presentation (e.g. printing), rather than computation. In the following example taken from Wickham, the predictor variable (treatment) is stored in the column headers rather than the actual data. As a result, these values are not included in the \JSON:

\begin{knitrout}
\definecolor{shadecolor}{rgb}{0.969, 0.969, 0.969}\color{fgcolor}\begin{kframe}
\begin{alltt}
\hlstd{x} \hlkwb{<-} \hlkwd{matrix}\hlstd{(}\hlkwd{c}\hlstd{(}\hlnum{NA}\hlstd{,} \hlnum{1}\hlstd{,} \hlnum{2}\hlstd{,} \hlnum{5}\hlstd{,} \hlnum{NA}\hlstd{,} \hlnum{3}\hlstd{),} \hlkwc{nrow} \hlstd{=} \hlnum{3}\hlstd{)}
\hlkwd{row.names}\hlstd{(x)} \hlkwb{<-} \hlkwd{c}\hlstd{(}\hlstr{"Joe"}\hlstd{,} \hlstr{"Jane"}\hlstd{,} \hlstr{"Mary"}\hlstd{)}
\hlkwd{colnames}\hlstd{(x)} \hlkwb{<-} \hlkwd{c}\hlstd{(}\hlstr{"Treatment A"}\hlstd{,} \hlstr{"Treatment B"}\hlstd{)}
\hlkwd{print}\hlstd{(x)}
\end{alltt}
\begin{verbatim}
     Treatment A Treatment B
Joe           NA           5
Jane           1          NA
Mary           2           3
\end{verbatim}
\begin{alltt}
\hlkwd{cat}\hlstd{(}\hlkwd{toJSON}\hlstd{(x))}
\end{alltt}
\begin{verbatim}
[ [ "NA", 5 ], [ 1, "NA" ], [ 2, 3 ] ]
\end{verbatim}
\end{kframe}
\end{knitrout}

Wickham recommends that the data be \emph{melted} into its \emph{tidy} form. Once the data is tidy, the \JSON encoding will naturally contain the treatment values:

\begin{knitrout}
\definecolor{shadecolor}{rgb}{0.969, 0.969, 0.969}\color{fgcolor}\begin{kframe}
\begin{alltt}
\hlkwd{library}\hlstd{(reshape2)}
\hlstd{y} \hlkwb{<-} \hlkwd{melt}\hlstd{(x,} \hlkwc{varnames} \hlstd{=} \hlkwd{c}\hlstd{(}\hlstr{"Subject"}\hlstd{,} \hlstr{"Treatment"}\hlstd{))}
\hlkwd{print}\hlstd{(y)}
\end{alltt}
\begin{verbatim}
  Subject   Treatment value
1     Joe Treatment A    NA
2    Jane Treatment A     1
3    Mary Treatment A     2
4     Joe Treatment B     5
5    Jane Treatment B    NA
6    Mary Treatment B     3
\end{verbatim}
\begin{alltt}
\hlkwd{cat}\hlstd{(}\hlkwd{toJSON}\hlstd{(y,} \hlkwc{pretty} \hlstd{=} \hlnum{TRUE}\hlstd{))}
\end{alltt}
\begin{verbatim}
[
  {
    "Subject" : "Joe",
    "Treatment" : "Treatment A"
  },
  {
    "Subject" : "Jane",
    "Treatment" : "Treatment A",
    "value" : 1
  },
  {
    "Subject" : "Mary",
    "Treatment" : "Treatment A",
    "value" : 2
  },
  {
    "Subject" : "Joe",
    "Treatment" : "Treatment B",
    "value" : 5
  },
  {
    "Subject" : "Jane",
    "Treatment" : "Treatment B"
  },
  {
    "Subject" : "Mary",
    "Treatment" : "Treatment B",
    "value" : 3
  }
]
\end{verbatim}
\end{kframe}
\end{knitrout}

In some other cases, the column headers actually do contain variable names, and melting is inappropriate. For data sets with records consisting of a set of named columns (fields), R has more natural and flexible class: the data-frame. The \toJSON method for data frames (described later) is more suitable when we want to refer to rows or fields by their name. Any matrix can easily be converted to a data-frame using the \texttt{as.data.frame} function:

\begin{knitrout}
\definecolor{shadecolor}{rgb}{0.969, 0.969, 0.969}\color{fgcolor}\begin{kframe}
\begin{alltt}
\hlkwd{cat}\hlstd{(}\hlkwd{toJSON}\hlstd{(}\hlkwd{as.data.frame}\hlstd{(x),} \hlkwc{pretty} \hlstd{=} \hlnum{TRUE}\hlstd{))}
\end{alltt}
\begin{verbatim}
[
  {
    "$row" : "Joe",
    "Treatment B" : 5
  },
  {
    "$row" : "Jane",
    "Treatment A" : 1
  },
  {
    "$row" : "Mary",
    "Treatment A" : 2,
    "Treatment B" : 3
  }
]
\end{verbatim}
\end{kframe}
\end{knitrout}

For some cases this results in the desired output, but in this example melting seems more appropriate.

\subsection{Lists}

The \texttt{list} is the most general purpose data structure in R. It holds an ordered set of elements, including other lists, each of arbitrary type and size. Two types of lists are distinguished: named lists and unnamed lists. A list is considered a named list if it has an attribute called \texttt{"names"}. In practice, a named list is any list for which we can access an element by its name, whereas elements of an unnamed lists can only be accessed using their index number:

\begin{knitrout}
\definecolor{shadecolor}{rgb}{0.969, 0.969, 0.969}\color{fgcolor}\begin{kframe}
\begin{alltt}
\hlstd{mylist1} \hlkwb{<-} \hlkwd{list}\hlstd{(}\hlkwc{foo} \hlstd{=} \hlnum{123}\hlstd{,} \hlkwc{bar} \hlstd{=} \hlnum{456}\hlstd{)}
\hlkwd{print}\hlstd{(mylist1}\hlopt{$}\hlstd{bar)}
\end{alltt}
\begin{verbatim}
[1] 456
\end{verbatim}
\begin{alltt}
\hlstd{mylist2} \hlkwb{<-} \hlkwd{list}\hlstd{(}\hlnum{123}\hlstd{,} \hlnum{456}\hlstd{)}
\hlkwd{print}\hlstd{(mylist2[[}\hlnum{2}\hlstd{]])}
\end{alltt}
\begin{verbatim}
[1] 456
\end{verbatim}
\end{kframe}
\end{knitrout}

\subsubsection{Unnamed lists}

Just like vectors, an unnamed list maps to a \JSON array:

\begin{knitrout}
\definecolor{shadecolor}{rgb}{0.969, 0.969, 0.969}\color{fgcolor}\begin{kframe}
\begin{alltt}
\hlkwd{cat}\hlstd{(}\hlkwd{toJSON}\hlstd{(}\hlkwd{list}\hlstd{(}\hlkwd{c}\hlstd{(}\hlnum{1}\hlstd{,} \hlnum{2}\hlstd{),} \hlstr{"test"}\hlstd{,} \hlnum{TRUE}\hlstd{,} \hlkwd{list}\hlstd{(}\hlkwd{c}\hlstd{(}\hlnum{1}\hlstd{,} \hlnum{2}\hlstd{)))))}
\end{alltt}
\begin{verbatim}
[ [ 1, 2 ], [ "test" ], [ true ], [ [ 1, 2 ] ] ]
\end{verbatim}
\end{kframe}
\end{knitrout}

Note that even though both vectors and lists are encoded using \JSON arrays, they can be distinguished from their contents: an R vector results in a \JSON array containing only primitives, whereas a list results in a \JSON array containing only objects and arrays. This allows the \JSON parser to reconstruct the original type from encoded vectors and arrays:

\begin{knitrout}
\definecolor{shadecolor}{rgb}{0.969, 0.969, 0.969}\color{fgcolor}\begin{kframe}
\begin{alltt}
\hlstd{x} \hlkwb{<-} \hlkwd{list}\hlstd{(}\hlkwd{c}\hlstd{(}\hlnum{1}\hlstd{,} \hlnum{2}\hlstd{,} \hlnum{NA}\hlstd{),} \hlstr{"test"}\hlstd{,} \hlnum{FALSE}\hlstd{,} \hlkwd{list}\hlstd{(}\hlkwc{foo} \hlstd{=} \hlstr{"bar"}\hlstd{))}
\hlkwd{identical}\hlstd{(}\hlkwd{fromJSON}\hlstd{(}\hlkwd{toJSON}\hlstd{(x)), x)}
\end{alltt}
\begin{verbatim}
[1] TRUE
\end{verbatim}
\end{kframe}
\end{knitrout}

 The only exception is the empty list and empty vector, which are both encoded as \texttt{[ ]} and therefore indistinguishable, but this is rarely a problem in practice. 

\subsubsection{Named lists}

A named list in R maps to a \JSON \emph{object}:

\begin{knitrout}
\definecolor{shadecolor}{rgb}{0.969, 0.969, 0.969}\color{fgcolor}\begin{kframe}
\begin{alltt}
\hlkwd{cat}\hlstd{(}\hlkwd{toJSON}\hlstd{(}\hlkwd{list}\hlstd{(}\hlkwc{foo} \hlstd{=} \hlkwd{c}\hlstd{(}\hlnum{1}\hlstd{,} \hlnum{2}\hlstd{),} \hlkwc{bar} \hlstd{=} \hlstr{"test"}\hlstd{)))}
\end{alltt}
\begin{verbatim}
{ "foo" : [ 1, 2 ], "bar" : [ "test" ] }
\end{verbatim}
\end{kframe}
\end{knitrout}

 Because a list can contain other lists, this works recursively:

\begin{knitrout}
\definecolor{shadecolor}{rgb}{0.969, 0.969, 0.969}\color{fgcolor}\begin{kframe}
\begin{alltt}
\hlkwd{cat}\hlstd{(}\hlkwd{toJSON}\hlstd{(}\hlkwd{list}\hlstd{(}\hlkwc{foo}\hlstd{=}\hlkwd{list}\hlstd{(}\hlkwc{bar}\hlstd{=}\hlkwd{list}\hlstd{(}\hlkwc{baz}\hlstd{=pi)))))}
\end{alltt}
\begin{verbatim}
{ "foo" : { "bar" : { "baz" : [ 3.14 ] } } }
\end{verbatim}
\end{kframe}
\end{knitrout}

 Named lists map almost perfectly to \JSON objects with one exception: list elements can have empty names:

\begin{knitrout}
\definecolor{shadecolor}{rgb}{0.969, 0.969, 0.969}\color{fgcolor}\begin{kframe}
\begin{alltt}
\hlstd{x} \hlkwb{<-} \hlkwd{list}\hlstd{(}\hlkwc{foo} \hlstd{=} \hlnum{123}\hlstd{,} \hlstr{"test"}\hlstd{,} \hlnum{TRUE}\hlstd{)}
\hlkwd{attr}\hlstd{(x,} \hlstr{"names"}\hlstd{)}
\end{alltt}
\begin{verbatim}
[1] "foo" ""    ""   
\end{verbatim}
\begin{alltt}
\hlstd{x}\hlopt{$}\hlstd{foo}
\end{alltt}
\begin{verbatim}
[1] 123
\end{verbatim}
\begin{alltt}
\hlstd{x[[}\hlnum{2}\hlstd{]]}
\end{alltt}
\begin{verbatim}
[1] "test"
\end{verbatim}
\end{kframe}
\end{knitrout}

 In a \JSON object, each element in an object must have a valid name. To ensure this property, \jsonlite uses the same solution as the \texttt{print} method, which is to fall back on indices for elements that do not have a proper name:

\begin{knitrout}
\definecolor{shadecolor}{rgb}{0.969, 0.969, 0.969}\color{fgcolor}\begin{kframe}
\begin{alltt}
\hlstd{x} \hlkwb{<-} \hlkwd{list}\hlstd{(}\hlkwc{foo} \hlstd{=} \hlnum{123}\hlstd{,} \hlstr{"test"}\hlstd{,} \hlnum{TRUE}\hlstd{)}
\hlkwd{print}\hlstd{(x)}
\end{alltt}
\begin{verbatim}
$foo
[1] 123

[[2]]
[1] "test"

[[3]]
[1] TRUE
\end{verbatim}
\begin{alltt}
\hlkwd{cat}\hlstd{(}\hlkwd{toJSON}\hlstd{(x))}
\end{alltt}
\begin{verbatim}
{ "foo" : [ 123 ], "2" : [ "test" ], "3" : [ true ] }
\end{verbatim}
\end{kframe}
\end{knitrout}

 This behavior ensures that all generated \JSON is valid, however named lists with empty names should be avoided where possible. When actually designing R objects that should be interoperable, it is recommended that each list element is given a proper name.

\subsection{Data frame}

The \texttt{data frame} is perhaps the most central data structure in R from the user point of view. This class holds tabular data in which each column is named and (usually) homogeneous. Conceptually it is very similar to a table in relational data bases such as \texttt{MySQL}, where \emph{fields} are referred to as \emph{column names}, and \emph{records} are called \emph{row names}. Like a matrix, a data frame can be subsetted with two indices, to extract certain rows and columns of the data:

\begin{knitrout}
\definecolor{shadecolor}{rgb}{0.969, 0.969, 0.969}\color{fgcolor}\begin{kframe}
\begin{alltt}
\hlkwd{is}\hlstd{(iris)}
\end{alltt}
\begin{verbatim}
[1] "data.frame" "list"       "oldClass"   "vector"    
\end{verbatim}
\begin{alltt}
\hlkwd{names}\hlstd{(iris)}
\end{alltt}
\begin{verbatim}
[1] "Sepal.Length" "Sepal.Width"  "Petal.Length" "Petal.Width" 
[5] "Species"     
\end{verbatim}
\begin{alltt}
\hlkwd{print}\hlstd{(iris[}\hlnum{1}\hlopt{:}\hlnum{3}\hlstd{,} \hlkwd{c}\hlstd{(}\hlnum{1}\hlstd{,} \hlnum{5}\hlstd{)])}
\end{alltt}
\begin{verbatim}
  Sepal.Length Species
1          5.1  setosa
2          4.9  setosa
3          4.7  setosa
\end{verbatim}
\begin{alltt}
\hlkwd{print}\hlstd{(iris[}\hlnum{1}\hlopt{:}\hlnum{3}\hlstd{,} \hlkwd{c}\hlstd{(}\hlstr{"Sepal.Width"}\hlstd{,} \hlstr{"Species"}\hlstd{)])}
\end{alltt}
\begin{verbatim}
  Sepal.Width Species
1         3.5  setosa
2         3.0  setosa
3         3.2  setosa
\end{verbatim}
\end{kframe}
\end{knitrout}

 For the previously discussed classes such as vectors and matrices, behavior of \jsonlite is quite similar to the other available packages that implement \texttt{toJSON} and \texttt{toJSON} functions, with only minor differences for missing values and edge cases. But when it comes to data frames, \jsonlite takes a completely different approach. The behavior of \jsonlite is designed for compatibility with conventional ways of encoding table-like structures outside the R community. The implementation is more complex, but results in a powerful and more natural way of interfacing data frames through \JSON and vice versa.

\subsubsection{Column based versus row based tables}

Generally speaking, tabular data structures can be implemented in two different ways: in a column based, or row based fashion. A column based structure consists of a named collection of equal-length, homogeneous arrays representing the table columns. In a row-based structure on the other hand, the table is implemented as a set of heterogeneous associative arrays representing table rows with field values for each particular record. Even though most languages provide flexible and abstracted interfaces that hide such implementation details from the user, they can have huge implications for performance. A column based structure is efficient for inserting or extracting certain columns of the data, but it is inefficient for manipulating individual rows. For example to insert a single row somewhere in the middle, each of the columns has to be sliced and stitched back together. For row-based implementations, it is the exact other way around: we can easily manipulate a particular record, but to insert/extract a whole column we would need to iterate over all records in the table and read/modify the appropriate field in each of them.

The data frame in R is implemented in a column based fashion: it constitutes of a \texttt{named list} of equal-length vectors. Thereby the columns in the data frame naturally inherit the properties from atomic vectors discussed before, such as homogeneity, missing values, etc. Another argument for column-based implementation is that statistical methods generally operate on columns. For example, the \texttt{lm} function fits a \emph{linear regression} by extracting the columns from a data frame as specified by the \texttt{formula} argument. R simply binds the specified columns together into a matrix \texttt{X} and calls out to a highly optimized FORTRAN subroutine to calculate the OLS estimates $\hat{\beta} = (X^TX)X^Ty$ using the $QR$ factorization of $X$. Many other statistical modeling functions follow similar steps, and are computationally efficient because of the column-based data storage in R. However, unfortunately R is an exception in its preference for column-based storage: most  languages, systems, databases, APIs, etc, are optimized for record based operations. For this reason, the conventional way to store and communicate tabular data in \JSON seems to almost exclusively row based. This discrepancy presents various complications when converting between data frames and \JSON. The remaining of this section discusses details and challenges of consistently mapping record based \JSON data as frequently encountered on the web, into column-based data frames which are convenient for statistical computing.

\subsubsection{Row based data frame encoding}

The encoding of data frames is one of the major differences between \jsonlite and implementations from other currently available packages. Instead of using the column-based encoding also used for lists, \jsonlite maps data frames by default to an array of records:

\begin{knitrout}
\definecolor{shadecolor}{rgb}{0.969, 0.969, 0.969}\color{fgcolor}\begin{kframe}
\begin{alltt}
\hlkwd{cat}\hlstd{(}\hlkwd{toJSON}\hlstd{(iris[}\hlnum{1}\hlopt{:}\hlnum{2}\hlstd{, ],} \hlkwc{pretty} \hlstd{=} \hlnum{TRUE}\hlstd{))}
\end{alltt}
\begin{verbatim}
[
  {
    "Sepal.Length" : 5.1,
    "Sepal.Width" : 3.5,
    "Petal.Length" : 1.4,
    "Petal.Width" : 0.2,
    "Species" : "setosa"
  },
  {
    "Sepal.Length" : 4.9,
    "Sepal.Width" : 3,
    "Petal.Length" : 1.4,
    "Petal.Width" : 0.2,
    "Species" : "setosa"
  }
]
\end{verbatim}
\end{kframe}
\end{knitrout}

 This output looks a bit like a list of named lists. However, there is one major difference: the individual records contain \JSON primitives, whereas lists always contain \JSON objects or arrays:

\begin{knitrout}
\definecolor{shadecolor}{rgb}{0.969, 0.969, 0.969}\color{fgcolor}\begin{kframe}
\begin{alltt}
\hlkwd{cat}\hlstd{(}\hlkwd{toJSON}\hlstd{(}\hlkwd{list}\hlstd{(}\hlkwd{list}\hlstd{(}\hlkwc{Species} \hlstd{=} \hlstr{"Foo"}\hlstd{,} \hlkwc{Width} \hlstd{=} \hlnum{21}\hlstd{)),} \hlkwc{pretty} \hlstd{=} \hlnum{TRUE}\hlstd{))}
\end{alltt}
\begin{verbatim}
[
  {
    "Species" : [
      "Foo"
    ],
    "Width" : [
      21
    ]
  }
]
\end{verbatim}
\end{kframe}
\end{knitrout}

 This leads to the following convention: when encoding R objects, \JSON primitives only appear in vectors and data-frame rows. Primitive values within a \JSON array indicate a vector, and \JSON primitives appearing inside a \JSON object indicate a data-frame row. A \JSON encoded \texttt{list}, (named or unnamed) will never contain \JSON primitives. This is a subtle but important convention that helps to distinguish between R classes from their \JSON representation, without explicitly encoding any metadata.

\subsubsection{Missing values in data frames}

The section on atomic vectors discussed two methods of encoding missing data appearing in a vector: either using strings or using the \JSON \texttt{null} type. When a missing value appears in a data frame, there is a third option: simply not include this field in \JSON record:

\begin{knitrout}
\definecolor{shadecolor}{rgb}{0.969, 0.969, 0.969}\color{fgcolor}\begin{kframe}
\begin{alltt}
\hlstd{x} \hlkwb{<-} \hlkwd{data.frame}\hlstd{(}\hlkwc{foo} \hlstd{=} \hlkwd{c}\hlstd{(}\hlnum{FALSE}\hlstd{,} \hlnum{TRUE}\hlstd{,} \hlnum{NA}\hlstd{,} \hlnum{NA}\hlstd{),} \hlkwc{bar} \hlstd{=} \hlkwd{c}\hlstd{(}\hlstr{"Aladdin"}\hlstd{,} \hlnum{NA}\hlstd{,} \hlnum{NA}\hlstd{,} \hlstr{"Mario"}\hlstd{))}
\hlkwd{print}\hlstd{(x)}
\end{alltt}
\begin{verbatim}
    foo     bar
1 FALSE Aladdin
2  TRUE    <NA>
3    NA    <NA>
4    NA   Mario
\end{verbatim}
\begin{alltt}
\hlkwd{cat}\hlstd{(}\hlkwd{toJSON}\hlstd{(x,} \hlkwc{pretty} \hlstd{=} \hlnum{TRUE}\hlstd{))}
\end{alltt}
\begin{verbatim}
[
  {
    "foo" : false,
    "bar" : "Aladdin"
  },
  {
    "foo" : true
  },
  {},
  {
    "bar" : "Mario"
  }
]
\end{verbatim}
\end{kframe}
\end{knitrout}

 The default behavior of \jsonlite is to omit missing data from records in a data frame. This seems to be the most conventional method used on the web, and we expect this encoding will most likely lead to the correct interpretation of \emph{missingness}, even in languages with no explicit notion of \texttt{NA}.

\subsubsection{Relational data: nested records}

Nested datasets are somewhat unusual in R, but frequently encountered in \JSON. Such structures do not really fit the vector based paradigm which makes them harder to manipulate in R. However, nested structures are too common in \JSON to ignore, and with a little work most cases still map to a data frame quite nicely. The most common scenario is a dataset in which a certain field within each record contains a \emph{subrecord} with additional fields. The \jsonlite implementation maps these subrecords to a nested data frame. Whereas the data frame class usually consists of vectors, technically a column can also be list or another data frame with matching dimension (this stretches the meaning of the word ``column'' a bit):

\begin{knitrout}
\definecolor{shadecolor}{rgb}{0.969, 0.969, 0.969}\color{fgcolor}\begin{kframe}
\begin{alltt}
\hlkwd{options}\hlstd{(}\hlkwc{stringsAsFactors}\hlstd{=}\hlnum{FALSE}\hlstd{)}
\hlstd{x} \hlkwb{<-} \hlkwd{data.frame}\hlstd{(}\hlkwc{driver} \hlstd{=} \hlkwd{c}\hlstd{(}\hlstr{"Bowser"}\hlstd{,} \hlstr{"Peach"}\hlstd{),} \hlkwc{occupation} \hlstd{=} \hlkwd{c}\hlstd{(}\hlstr{"Koopa"}\hlstd{,} \hlstr{"Princess"}\hlstd{))}
\hlstd{x}\hlopt{$}\hlstd{vehicle} \hlkwb{<-} \hlkwd{data.frame}\hlstd{(}\hlkwc{model} \hlstd{=} \hlkwd{c}\hlstd{(}\hlstr{"Piranha Prowler"}\hlstd{,} \hlstr{"Royal Racer"}\hlstd{))}
\hlstd{x}\hlopt{$}\hlstd{vehicle}\hlopt{$}\hlstd{stats} \hlkwb{<-} \hlkwd{data.frame}\hlstd{(}\hlkwc{speed} \hlstd{=} \hlkwd{c}\hlstd{(}\hlnum{55}\hlstd{,} \hlnum{34}\hlstd{),} \hlkwc{weight} \hlstd{=} \hlkwd{c}\hlstd{(}\hlnum{67}\hlstd{,} \hlnum{24}\hlstd{),} \hlkwc{drift} \hlstd{=} \hlkwd{c}\hlstd{(}\hlnum{35}\hlstd{,} \hlnum{32}\hlstd{))}
\hlkwd{str}\hlstd{(x)}
\end{alltt}
\begin{verbatim}
'data.frame':	2 obs. of  3 variables:
 $ driver    : chr  "Bowser" "Peach"
 $ occupation: chr  "Koopa" "Princess"
 $ vehicle   :'data.frame':	2 obs. of  2 variables:
  ..$ model: chr  "Piranha Prowler" "Royal Racer"
  ..$ stats:'data.frame':	2 obs. of  3 variables:
  .. ..$ speed : num  55 34
  .. ..$ weight: num  67 24
  .. ..$ drift : num  35 32
\end{verbatim}
\begin{alltt}
\hlkwd{cat}\hlstd{(}\hlkwd{toJSON}\hlstd{(x,} \hlkwc{pretty}\hlstd{=}\hlnum{TRUE}\hlstd{))}
\end{alltt}
\begin{verbatim}
[
  {
    "driver" : "Bowser",
    "occupation" : "Koopa",
    "vehicle" : {
      "model" : "Piranha Prowler",
      "stats" : {
        "speed" : 55,
        "weight" : 67,
        "drift" : 35
      }
    }
  },
  {
    "driver" : "Peach",
    "occupation" : "Princess",
    "vehicle" : {
      "model" : "Royal Racer",
      "stats" : {
        "speed" : 34,
        "weight" : 24,
        "drift" : 32
      }
    }
  }
]
\end{verbatim}
\begin{alltt}
\hlstd{myjson} \hlkwb{<-} \hlkwd{toJSON}\hlstd{(x)}
\hlstd{y} \hlkwb{<-} \hlkwd{fromJSON}\hlstd{(myjson)}
\hlkwd{identical}\hlstd{(x,y)}
\end{alltt}
\begin{verbatim}
[1] TRUE
\end{verbatim}
\end{kframe}
\end{knitrout}

 When encountering \JSON data containing nested records on the web, chances are that these data were generated from \emph{relational} database. The \JSON field containing a subrecord represents a \emph{foreign key} pointing to a record in an external table. For the purpose of encoding these into a single \JSON structure, the tables were joined into a nested structure. The directly nested subrecord represents a \emph{one-to-one} or \emph{many-to-one} relation between the parent and child table, and is most naturally stored in R using a nested data frame. In the example above, the \texttt{vehicle} field points to a table of vehicles, which in turn contains a \texttt{stats} field pointing to a table of stats. When there is no more than one subrecord for each record, we easily \emph{flatten} the structure into a single non-nested data frame.

\begin{knitrout}
\definecolor{shadecolor}{rgb}{0.969, 0.969, 0.969}\color{fgcolor}\begin{kframe}
\begin{alltt}
\hlstd{z} \hlkwb{<-} \hlkwd{cbind}\hlstd{(x[}\hlkwd{c}\hlstd{(}\hlstr{"driver"}\hlstd{,} \hlstr{"occupation"}\hlstd{)], x}\hlopt{$}\hlstd{vehicle[}\hlstr{"model"}\hlstd{], x}\hlopt{$}\hlstd{vehicle}\hlopt{$}\hlstd{stats)}
\hlkwd{str}\hlstd{(z)}
\end{alltt}
\begin{verbatim}
'data.frame':	2 obs. of  6 variables:
 $ driver    : chr  "Bowser" "Peach"
 $ occupation: chr  "Koopa" "Princess"
 $ model     : chr  "Piranha Prowler" "Royal Racer"
 $ speed     : num  55 34
 $ weight    : num  67 24
 $ drift     : num  35 32
\end{verbatim}
\end{kframe}
\end{knitrout}

\subsubsection{Relational data: nested tables}

The one-to-one relation discussed above is relatively easy to store in R, because each record contains at most one subrecord. Therefore we can use either a nested data frame, or flatten the data frame. However, things get more difficult when \JSON records contain a field with a nested array. Such a structure appears in relational data in case of a \emph{one-to-many} relation. A standard textbook illustration is the relation between authors and titles. For example, a field can contain an array of values:

\begin{knitrout}
\definecolor{shadecolor}{rgb}{0.969, 0.969, 0.969}\color{fgcolor}\begin{kframe}
\begin{alltt}
\hlstd{x} \hlkwb{<-} \hlkwd{data.frame}\hlstd{(}\hlkwc{author} \hlstd{=} \hlkwd{c}\hlstd{(}\hlstr{"Homer"}\hlstd{,} \hlstr{"Virgil"}\hlstd{,} \hlstr{"Jeroen"}\hlstd{))}
\hlstd{x}\hlopt{$}\hlstd{poems} \hlkwb{<-} \hlkwd{list}\hlstd{(}\hlkwd{c}\hlstd{(}\hlstr{"Iliad"}\hlstd{,} \hlstr{"Odyssey"}\hlstd{),} \hlkwd{c}\hlstd{(}\hlstr{"Eclogues"}\hlstd{,} \hlstr{"Georgics"}\hlstd{,} \hlstr{"Aeneid"}\hlstd{),} \hlkwd{vector}\hlstd{());}
\hlkwd{names}\hlstd{(x)}
\end{alltt}
\begin{verbatim}
[1] "author" "poems" 
\end{verbatim}
\begin{alltt}
\hlkwd{cat}\hlstd{(}\hlkwd{toJSON}\hlstd{(x,} \hlkwc{pretty} \hlstd{=} \hlnum{TRUE}\hlstd{))}
\end{alltt}
\begin{verbatim}
[
  {
    "author" : "Homer",
    "poems" : [
      "Iliad",
      "Odyssey"
    ]
  },
  {
    "author" : "Virgil",
    "poems" : [
      "Eclogues",
      "Georgics",
      "Aeneid"
    ]
  },
  {
    "author" : "Jeroen",
    "poems" : []
  }
]
\end{verbatim}
\end{kframe}
\end{knitrout}

 As can be seen from the example, the way to store this in a data frame is using a list of character vectors. This works, and although unconventional, we can still create and read such structures in R relatively easily. However, in practice the one-to-many relation is often more complex. It results in fields containing a \emph{set of records}. In R, the only way to model this is as a column containing a list of data frames, one separate data frame for each row:

\begin{knitrout}
\definecolor{shadecolor}{rgb}{0.969, 0.969, 0.969}\color{fgcolor}\begin{kframe}
\begin{alltt}
\hlstd{x} \hlkwb{<-} \hlkwd{data.frame}\hlstd{(}\hlkwc{author} \hlstd{=} \hlkwd{c}\hlstd{(}\hlstr{"Homer"}\hlstd{,} \hlstr{"Virgil"}\hlstd{,} \hlstr{"Jeroen"}\hlstd{))}
\hlstd{x}\hlopt{$}\hlstd{poems} \hlkwb{<-} \hlkwd{list}\hlstd{(}
  \hlkwd{data.frame}\hlstd{(}\hlkwc{title}\hlstd{=}\hlkwd{c}\hlstd{(}\hlstr{"Iliad"}\hlstd{,} \hlstr{"Odyssey"}\hlstd{),} \hlkwc{year}\hlstd{=}\hlkwd{c}\hlstd{(}\hlopt{-}\hlnum{1194}\hlstd{,} \hlopt{-}\hlnum{800}\hlstd{)),}
  \hlkwd{data.frame}\hlstd{(}\hlkwc{title}\hlstd{=}\hlkwd{c}\hlstd{(}\hlstr{"Eclogues"}\hlstd{,} \hlstr{"Georgics"}\hlstd{,} \hlstr{"Aeneid"}\hlstd{),} \hlkwc{year}\hlstd{=}\hlkwd{c}\hlstd{(}\hlopt{-}\hlnum{44}\hlstd{,} \hlopt{-}\hlnum{29}\hlstd{,} \hlopt{-}\hlnum{19}\hlstd{)),}
  \hlkwd{data.frame}\hlstd{()}
\hlstd{)}
\hlkwd{cat}\hlstd{(}\hlkwd{toJSON}\hlstd{(x,} \hlkwc{pretty}\hlstd{=}\hlnum{TRUE}\hlstd{))}
\end{alltt}
\begin{verbatim}
[
  {
    "author" : "Homer",
    "poems" : [
      {
        "title" : "Iliad",
        "year" : -1194
      },
      {
        "title" : "Odyssey",
        "year" : -800
      }
    ]
  },
  {
    "author" : "Virgil",
    "poems" : [
      {
        "title" : "Eclogues",
        "year" : -44
      },
      {
        "title" : "Georgics",
        "year" : -29
      },
      {
        "title" : "Aeneid",
        "year" : -19
      }
    ]
  },
  {
    "author" : "Jeroen",
    "poems" : []
  }
]
\end{verbatim}
\end{kframe}
\end{knitrout}

 Because R doesn't have native support for relational data, there is no natural class to store such structures. The best we can do is a column containing a list of sub-dataframes. This does the job, and allows the R user to access or generate nested \JSON structures. However, a data frame like this cannot be flattened, and the class does not guarantee that each of the individual nested data frames contain the same fields, as would be the case in an actual relational data base.

\section{Structural consistency and type safety in dynamic data}

Systems that automatically exchange information over some interface, protocol or API require well defined and unambiguous meaning and arrangement of data. In order to process and interpret input and output, contents must obey a steady structure. Such structures are usually described either informally in documentation or more formally in a schema language. The previous section emphasized the importance of consistency in the mapping between \JSON data and R classes. This section takes a higher level view and explains the importance of structure consistency for dynamic data. This topic can be a bit subtle because it refers to consistency among different instantiations of a \JSON structure, rather than a single case. We try to clarify by breaking down the concept into two important parts, and illustrate with analogies and examples from R.

\subsection{Classes, types and data}

Most object-oriented languages are designed with the idea that all objects of a certain class implement the same fields and methods. In strong-typed languages such as \texttt{S4} or \texttt{Java}, names and types of the fields are formally declared in a class definition. In other languages such as \texttt{S3} or \texttt{JavaScript}, the fields are not enforced by the language but rather at the discretion of the programmer. However one way or another they all assume that members of a certain class agree on field names and types, so that the same methods can be applied to any object of a particular class. This basic principle holds for dynamic data exactly the same way as for objects. Software that process dynamic data can only work reliably if the various elements of the data have consistent names and structure. Consensus must exist between the different parties on data that is exchanged as part an interface or protocol. This requires the structure to follow some sort of template that specifies which attributes can appear in the data, what they mean and how they are composed. Thereby each possible scenario can be accounted for in the software so that data gets interpreted/processed appropriately and no exceptions arise during run-time. 

Some data interchange formats such as \texttt{XML} or \texttt{Protocol Buffers} take a formal approach to this matter, and have well established \emph{schema languages} and \emph{interface description languages}. Using such a meta language it is possible to define the exact structure, properties and actions of data interchange in a formal arrangement. However, in \JSON, such formal definitions are relatively uncommon. Some initiatives for \JSON schema languages exist \citep{jsonschema}, but they are not very well established and rarely seen in practice. One reason for this might be that defining and implementing formal schemas is complicated and a lot of work which defeats the purpose of using an lightweight format such as \JSON in the first place. But another reason is that it is often simply not necessary to be overly formal. The \JSON format is simple and intuitive, and under some general conventions, a well chosen example can suffice to characterize the structure. This section describes two important rules that are required to ensure that data exchange using \JSON is type safe.

\subsection{Rule 1: Fixed keys}

When using \JSON without a schema, there are no restrictions on the keys (field names) that can appear in a particular object. However, an API that returns a different set of keys every time it is called makes it very difficult to write software to process these data. Hence, the first rule is to limit \JSON interfaces to a finite set of keys that are known \emph{a priory} by all parties. It can be helpful to think about this in analogy with for example a relational database. Here, the database model separates the data from metadata. At run time, records can be inserted or deleted, and a certain query might return different data each time it is executed. But for a given query, each execution will return exactly the same \emph{field names}; hence as long as the table definitions are unchanged, the \emph{structure} of the output consistent. Client software needs this structure to validate input, optimize implementation, and process each part of the data appropriately. In \JSON, data and metadata are not formally separated as in a database, but similar principles that hold for fields in a database, apply to keys in dynamic \JSON data.  
 
A beautiful example of this in practice was given by Mike Dewar at the New York Open Statistical Programming Meetup on Jan. 12, 2012 \citep{jsonkeys}. In his talk he emphasizes to use \JSON keys only for \emph{names}, and not for \emph{data}. He refers to this principle as the ``golden rule'', and explains how he learned his lesson the hard way. In one of his early applications, timeseries data was encoded by using the epoch timestamp as the \JSON key. Therefore the keys are different each time the query is executed:

\begin{verbatim}
[
  { "1325344443" : 124 },
  { "1325344456" : 131 },
  { "1325344478" : 137 },
]
\end{verbatim}

 Even though being valid \JSON, dynamic keys as in the example above are likely to introduce trouble. Most software will have great difficulty processing these values if we can not specify the keys in the code. Moreover when documenting the API, either informally or formally using a schema language, we need to describe for each property in the data what the value means and is composed of. Thereby a client or consumer can implement code that interprets and process each element in the data in an appropriate manner. Both the documentation and interpretation of \JSON data rely on fixed keys with well defined meaning. Also note that the structure is difficult to extend in the future. If we want to add an additional property to each observation, the entire structure needs to change. In his talk, Dewar explains that life gets much easier when we switch to the following encoding:

\begin{verbatim}
[
  { "time": "1325344443" : "price": 124 },
  { "time": "1325344456" : "price": 131 },
  { "time": "1325344478" : "price": 137 }
]
\end{verbatim}

 This structure will play much nicer with existing software that assumes fixed keys. Moreover, the structure can easily be described in documentation, or captured in a schema. Even when we have no intention of writing documentation or a schema for a dynamic \JSON source, it is still wise to design the structure in such away that it \emph{could} be described by a schema. When the keys are fixed, a well chosen example can provide all the information required for the consumer to implement client code. Also note that the new structure is extensible: additional properties can be added to each observation without breaking backward compatibility.  

In the context of R, consistency of keys is closely related to Wikcham's concept of \emph{tidy data} discussed earlier. Wickham states that the most common reason for messy data are column headers containing values instead of variable names. Column headers in tabular datasets become keys when converted to \JSON. Therefore, when headers are actually values, \JSON keys contain in fact data and can become unpredictable. The cure to inconsistent keys is almost always to tidy the data according to recommendations given by \cite{tidydata}. 
 
\subsection{Rule 2: Consistent types}

In a strong typed language, fields declare their class before any values are assigned. Thereby the type of a given field is identical in all objects of a particular class, and arrays only contain objects of a single type. The \texttt{S3} system in R is weakly typed and puts no formal restrictions on the class of a certain properties, or the types of objects that can be combined into a collection. For example, the list below contains a character vector, a numeric vector and a list:

\begin{knitrout}
\definecolor{shadecolor}{rgb}{0.969, 0.969, 0.969}\color{fgcolor}\begin{kframe}
\begin{alltt}
\hlcom{# Heterogeneous lists are bad!}
\hlstd{x} \hlkwb{<-} \hlkwd{list}\hlstd{(}\hlstr{"FOO"}\hlstd{,} \hlnum{1}\hlopt{:}\hlnum{3}\hlstd{,} \hlkwd{list}\hlstd{(}\hlkwc{bar} \hlstd{= pi))}
\hlkwd{cat}\hlstd{(}\hlkwd{toJSON}\hlstd{(x))}
\end{alltt}
\begin{verbatim}
[ [ "FOO" ], [ 1, 2, 3 ], { "bar" : [ 3.14 ] } ]
\end{verbatim}
\end{kframe}
\end{knitrout}

 However even though it is possible to generate such \JSON, it is bad practice. Fields or collections with ambiguous object types are difficult to describe, interpret and process in the context of inter-system communication. When using \JSON to exchange dynamic data, it is important that each property and array is \emph{type consistent}. In dynamically typed languages, the programmer needs to make sure that properties are of the correct type before encoding into \JSON. It also means that \texttt{unnamed lists} in R should generally be avoided when designing interoperable structures because this type is not homogeneous. 
 
 Note that consistency is somewhat subjective as it refers to the \emph{meaning} of the elements; they do not necessarily have precisely the same structure. What is important is to keep in mind that the consumer of the data can interpret and process each element identically, e.g. iterate over the elements in the collection and apply the same method to each of them. To illustrate this, lets take the example of the data frame: 

\begin{knitrout}
\definecolor{shadecolor}{rgb}{0.969, 0.969, 0.969}\color{fgcolor}\begin{kframe}
\begin{alltt}
\hlcom{# conceptually homogenous array}
\hlstd{x} \hlkwb{<-} \hlkwd{data.frame}\hlstd{(}\hlkwc{name} \hlstd{=} \hlkwd{c}\hlstd{(}\hlstr{"Jay"}\hlstd{,} \hlstr{"Mary"}\hlstd{,} \hlnum{NA}\hlstd{,} \hlnum{NA}\hlstd{),} \hlkwc{gender} \hlstd{=} \hlkwd{c}\hlstd{(}\hlstr{"M"}\hlstd{,} \hlnum{NA}\hlstd{,} \hlnum{NA}\hlstd{,} \hlstr{"F"}\hlstd{))}
\hlkwd{cat}\hlstd{(}\hlkwd{toJSON}\hlstd{(x,} \hlkwc{pretty} \hlstd{=} \hlnum{TRUE}\hlstd{))}
\end{alltt}
\begin{verbatim}
[
  {
    "name" : "Jay",
    "gender" : "M"
  },
  {
    "name" : "Mary"
  },
  {},
  {
    "gender" : "F"
  }
]
\end{verbatim}
\end{kframe}
\end{knitrout}

 The \JSON array above has 4 elements, each of which a \JSON object. However, due to the \texttt{NA} values, not all elements have an identical structure: some records have more fields than others. But as long as they are conceptually the same type (e.g. a person), the consumer can iterate over the elements to process each person in the set according to a predefined action. For example each element could be used to construct a \texttt{Person} object. A collection of different object classes should be separated and organized using a named list:

\begin{knitrout}
\definecolor{shadecolor}{rgb}{0.969, 0.969, 0.969}\color{fgcolor}\begin{kframe}
\begin{alltt}
\hlstd{x} \hlkwb{<-} \hlkwd{list}\hlstd{(}
  \hlkwc{humans} \hlstd{=} \hlkwd{data.frame}\hlstd{(}\hlkwc{name} \hlstd{=} \hlkwd{c}\hlstd{(}\hlstr{"Jay"}\hlstd{,} \hlstr{"Mary"}\hlstd{),} \hlkwc{married} \hlstd{=} \hlkwd{c}\hlstd{(}\hlnum{TRUE}\hlstd{,} \hlnum{FALSE}\hlstd{)),}
  \hlkwc{horses} \hlstd{=} \hlkwd{data.frame}\hlstd{(}\hlkwc{name} \hlstd{=} \hlkwd{c}\hlstd{(}\hlstr{"Star"}\hlstd{,} \hlstr{"Dakota"}\hlstd{),} \hlkwc{price} \hlstd{=} \hlkwd{c}\hlstd{(}\hlnum{5000}\hlstd{,} \hlnum{30000}\hlstd{))}
\hlstd{)}
\hlkwd{cat}\hlstd{(}\hlkwd{toJSON}\hlstd{(x,} \hlkwc{pretty}\hlstd{=}\hlnum{TRUE}\hlstd{))}
\end{alltt}
\begin{verbatim}
{
  "humans" : [
    {
      "name" : "Jay",
      "married" : true
    },
    {
      "name" : "Mary",
      "married" : false
    }
  ],
  "horses" : [
    {
      "name" : "Star",
      "price" : 5000
    },
    {
      "name" : "Dakota",
      "price" : 30000
    }
  ]
}
\end{verbatim}
\end{kframe}
\end{knitrout}

 This might seem obvious, but dynamic languages such as R can make it dangerously tempting to create data containing mixed-type properties or collections. We already mentioned the example of other \JSON packages which encode an atomic vector either as \JSON primitive or \JSON array, depending on its length. Such inconsistent typing is very difficult for clients to predict and a likely source of nasty bugs. Using consistent field names/types and homogeneous \JSON arrays is a strong convention among public \JSON APIs, for good reasons. We recommend R users to respect these conventions when generating \JSON data in R.

\begin{appendices}

\section{Public \JSON APIs}

This section lists some examples of public HTTP APIs that publish data in \JSON format. These are great to get a sense of the complex structures that are encountered in real world \JSON data. All services are free, but some require registration/authentication.

Each example returns lots of data, therefore output is ommitted in this document. The reader is encouraged to run the examples in R and inspect the output manually. 

\subsection{No authentication required}

The following APIs allow for (limited) use without any form of registration:

\subsubsection*{Github}

Github is an online code repository and has APIs to get live data on almost all activity. Below some examples from a well known R package and author:

\begin{knitrout}
\definecolor{shadecolor}{rgb}{0.969, 0.969, 0.969}\color{fgcolor}\begin{kframe}
\begin{alltt}
\hlstd{hadley_orgs} \hlkwb{<-} \hlkwd{fromJSON}\hlstd{(}\hlstr{"https://api.github.com/users/hadley/orgs"}\hlstd{)}
\hlstd{hadley_repos} \hlkwb{<-} \hlkwd{fromJSON}\hlstd{(}\hlstr{"https://api.github.com/users/hadley/repos"}\hlstd{)}
\hlstd{gg_issues} \hlkwb{<-} \hlkwd{fromJSON}\hlstd{(}\hlstr{"https://api.github.com/repos/hadley/ggplot2/issues"}\hlstd{)}
\hlstd{gg_commits} \hlkwb{<-} \hlkwd{fromJSON}\hlstd{(}\hlstr{"https://api.github.com/repos/hadley/ggplot2/commits"}\hlstd{)}
\end{alltt}
\end{kframe}
\end{knitrout}

\subsubsection*{CitiBike NYC}

A single public API that shows location, status and current availability for all stations in the New York City bike sharing initative. 

\begin{knitrout}
\definecolor{shadecolor}{rgb}{0.969, 0.969, 0.969}\color{fgcolor}\begin{kframe}
\begin{alltt}
\hlstd{citibike} \hlkwb{<-} \hlkwd{fromJSON}\hlstd{(}\hlstr{"http://citibikenyc.com/stations/json"}\hlstd{)}
\end{alltt}
\end{kframe}
\end{knitrout}

\subsubsection*{AngelList}

AngelList is a job listing directory for startups:

\begin{knitrout}
\definecolor{shadecolor}{rgb}{0.969, 0.969, 0.969}\color{fgcolor}\begin{kframe}
\begin{alltt}
\hlstd{res} \hlkwb{<-} \hlkwd{fromJSON}\hlstd{(}\hlstr{"http://api.angel.co/1/tags/59/startups"}\hlstd{)}
\hlstd{res}\hlopt{$}\hlstd{startups}
\end{alltt}
\end{kframe}
\end{knitrout}

\subsubsection*{Ergast}

The Ergast Developer API is an experimental web service which provides a historical record of motor racing data for non-commercial purposes.

\begin{knitrout}
\definecolor{shadecolor}{rgb}{0.969, 0.969, 0.969}\color{fgcolor}\begin{kframe}
\begin{alltt}
\hlstd{races} \hlkwb{<-} \hlkwd{fromJSON}\hlstd{(}\hlstr{"http://ergast.com/api/f1/2012/1/results.json"}\hlstd{)}
\hlstd{races}\hlopt{$}\hlstd{RaceTable}\hlopt{$}\hlstd{Races}\hlopt{$}\hlstd{MRData}\hlopt{$}\hlstd{Results[[}\hlnum{1}\hlstd{]]}\hlopt{$}\hlstd{Driver}
\end{alltt}
\end{kframe}
\end{knitrout}

\subsection{Free registration required}

The following APIs require (free) registration of a key/token. In each case, registration is easy and a key will be emailed. This key has to be appended to the requests to query the APIs. The code below includes some example keys for illustration purposes.

\subsubsection*{New York Times}

The New York Times has several free APIs that are part of the NYT developer network. These interface to data from various departments, such as news articles, book reviews, real estate, etc.
 
\begin{knitrout}
\definecolor{shadecolor}{rgb}{0.969, 0.969, 0.969}\color{fgcolor}\begin{kframe}
\begin{alltt}
\hlcom{# Register keys at http://developer.nytimes.com/docs/reference/keys}

\hlcom{# search for articles}
\hlstd{article_key} \hlkwb{=} \hlstr{"&api-key=c2fede7bd9aea57c898f538e5ec0a1ee:6:68700045"}
\hlstd{url} \hlkwb{=} \hlstr{"http://api.nytimes.com/svc/search/v2/articlesearch.json?q=obamacare+socialism"}
\hlstd{articles} \hlkwb{<-} \hlkwd{fromJSON}\hlstd{(}\hlkwd{paste0}\hlstd{(url, article_key))}

\hlcom{# search for best sellers}
\hlstd{bestseller_key} \hlkwb{=} \hlstr{"&api-key=5e260a86a6301f55546c83a47d139b0d:3:68700045"}
\hlstd{url} \hlkwb{=} \hlstr{"http://api.nytimes.com/svc/books/v2/lists/overview.json?published_date=2013-01-01"}
\hlstd{bestsellers} \hlkwb{<-} \hlkwd{fromJSON}\hlstd{(}\hlkwd{paste0}\hlstd{(url, bestseller_key))}

\hlcom{# movie reviews}
\hlstd{movie_key} \hlkwb{=} \hlstr{"&api-key=5a3daaeee6bbc6b9df16284bc575e5ba:0:68700045"}
\hlstd{url} \hlkwb{=} \hlstr{"http://api.nytimes.com/svc/movies/v2/reviews/dvd-picks.json?order=by-date"}
\hlstd{reviews} \hlkwb{<-} \hlkwd{fromJSON}\hlstd{(}\hlkwd{paste0}\hlstd{(url, movie_key))}
\end{alltt}
\end{kframe}
\end{knitrout}

\subsubsection*{CrunchBase}

CrunchBase is the free database of technology companies, people, and investors that anyone can edit.

\begin{knitrout}
\definecolor{shadecolor}{rgb}{0.969, 0.969, 0.969}\color{fgcolor}\begin{kframe}
\begin{alltt}
\hlstd{key} \hlkwb{<-} \hlstr{"f6dv6cas5vw7arn5b9d7mdm3"}
\hlstd{res} \hlkwb{<-} \hlkwd{fromJSON}\hlstd{(}\hlkwd{paste0}\hlstd{(}\hlstr{"http://api.crunchbase.com/v/1/search.js?query=R&api_key="}\hlstd{, key))}
\hlkwd{str}\hlstd{(res}\hlopt{$}\hlstd{results)}
\end{alltt}
\end{kframe}
\end{knitrout}

\subsubsection*{Sunlight Foundation}

The Sunlight Foundation is a non-profit that helps to make government transparent and accountable through data, tools, policy and journalism.

\begin{knitrout}
\definecolor{shadecolor}{rgb}{0.969, 0.969, 0.969}\color{fgcolor}\begin{kframe}
\begin{alltt}
\hlcom{#register key at http://sunlightfoundation.com/api/accounts/register/ }
\hlstd{key} \hlkwb{<-} \hlstr{"&apikey=39c83d5a4acc42be993ee637e2e4ba3d"}

\hlcom{#some queries}
\hlstd{drones} \hlkwb{<-} \hlkwd{fromJSON}\hlstd{(}\hlkwd{paste0}\hlstd{(}\hlstr{"http://openstates.org/api/v1/bills/?q=drone"}\hlstd{, key))}
\hlstd{word} \hlkwb{<-} \hlkwd{fromJSON}\hlstd{(}\hlkwd{paste0}\hlstd{(}\hlstr{"http://capitolwords.org/api/1/dates.json?phrase=obamacare"}\hlstd{, key))}
\hlstd{legislators} \hlkwb{<-} \hlkwd{fromJSON}\hlstd{(}\hlkwd{paste0}\hlstd{(}\hlstr{"http://congress.api.sunlightfoundation.com/"}\hlstd{,}
  \hlstr{"legislators/locate?latitude=42.96&longitude=-108.09"}\hlstd{, key))}
\end{alltt}
\end{kframe}
\end{knitrout}

\subsection{OAuth2 authentication} 
 
\subsubsection*{Twitter}

The twitter API requires a more elaborate authentication process based on the \texttt{OAuth2} protocol. Some example code:

\begin{knitrout}
\definecolor{shadecolor}{rgb}{0.969, 0.969, 0.969}\color{fgcolor}\begin{kframe}
\begin{alltt}
\hlcom{#Create your own appication key at https://dev.twitter.com/apps}
\hlstd{consumer_key} \hlkwb{=} \hlstr{"EZRy5JzOH2QQmVAe9B4j2w"}\hlstd{;}
\hlstd{consumer_secret} \hlkwb{=} \hlstr{"OIDC4MdfZJ82nbwpZfoUO4WOLTYjoRhpHRAWj6JMec"}\hlstd{;}

\hlcom{#basic auth}
\hlkwd{library}\hlstd{(httr)}
\hlstd{secret} \hlkwb{<-} \hlstd{RCurl::}\hlkwd{base64}\hlstd{(}\hlkwd{paste}\hlstd{(consumer_key, consumer_secret,} \hlkwc{sep}\hlstd{=}\hlstr{":"}\hlstd{));}
\hlstd{req} \hlkwb{<-} \hlkwd{POST}\hlstd{(}\hlstr{"https://api.twitter.com/oauth2/token"}\hlstd{,}
  \hlkwd{config}\hlstd{(}\hlkwc{httpheader} \hlstd{=} \hlkwd{c}\hlstd{(}
    \hlstr{"Authorization"} \hlstd{=} \hlkwd{paste}\hlstd{(}\hlstr{"Basic"}\hlstd{, secret),}
    \hlstr{"Content-Type"} \hlstd{=} \hlstr{"application/x-www-form-urlencoded;charset=UTF-8"}
  \hlstd{)),}
  \hlkwc{body} \hlstd{=} \hlstr{"grant_type=client_credentials"}\hlstd{,}
  \hlkwc{multipart} \hlstd{=} \hlnum{FALSE}
\hlstd{);}

\hlstd{res} \hlkwb{<-} \hlkwd{fromJSON}\hlstd{(}\hlkwd{rawToChar}\hlstd{(req}\hlopt{$}\hlstd{content))}
\hlstd{token} \hlkwb{<-} \hlkwd{paste}\hlstd{(}\hlstr{"Bearer"}\hlstd{, res}\hlopt{$}\hlstd{access_token);}

\hlcom{#Actual API call}
\hlstd{url} \hlkwb{=} \hlstr{"https://api.twitter.com/1.1/statuses/user_timeline.json?count=10&screen_name=UCLA"}
\hlstd{call1} \hlkwb{<-} \hlkwd{GET}\hlstd{(url,} \hlkwd{config}\hlstd{(}\hlkwc{httpheader} \hlstd{=} \hlkwd{c}\hlstd{(}\hlstr{"Authorization"} \hlstd{= token)))}
\hlstd{obj1} \hlkwb{<-} \hlkwd{fromJSON}\hlstd{(}\hlkwd{rawToChar}\hlstd{(call1}\hlopt{$}\hlstd{content))}
\end{alltt}
\end{kframe}
\end{knitrout}

\section{Simple \JSON RPC with OpenCPU}

The \jsonlite package is used by \texttt{OpenCPU} to convert between \JSON data and R ojects. Thereby clients can retrieve R objects, or remotely call R functions using \JSON where the function arguments as well as function return value are \JSON objects. For example to download the \texttt{Boston} data from the \texttt{MASS} package:\\

\begin{tabular}{|l|l|}
  \hline
     \textbf{Command in R} & \textbf{Example URL on OpenCPU} \\
  \hline
     \texttt{toJSON(Boston, digits=4)} & \url{https://cran.ocpu.io/MASS/data/Boston/json?digits=4} \\
  \hline 
     \texttt{toJSON(Boston, dataframe="col")} & \url{https://cran.ocpu.io/MASS/data/Boston/json?dataframe=col} \\
  \hline
     \texttt{toJSON(Boston, pretty=FALSE)} & \url{https://cran.ocpu.io/MASS/data/Boston/json?pretty=false} \\
  \hline  
\end{tabular}
\newline

To calculate the variance of some the numbers \texttt{1:9} in the command line using using \texttt{curl}:

\begin{Verbatim}[frame=single]
curl https://demo.ocpu.io/stats/R/var/json -d 'x=[1,2,3,4,5,6,7,8,9]'
\end{Verbatim}

Or equivalently post the entire body in \JSON format:

\begin{Verbatim}[frame=single]
curl https://demo.ocpu.io/stats/R/var/json -H "Content-Type: application/json" \
-d '{"x":[1,2,3,4,5,6,7,8,9]}'
\end{Verbatim}

Below an example where we call the \texttt{melt} function from the \texttt{reshape2} package using some example rows from the \texttt{airquality} data. Here both input and output consist of a data frame.

\begin{Verbatim}[frame=single]
curl https://cran.ocpu.io/reshape2/R/melt/json -d 'id=["Month", "Day"]&data=[ 
  { "Ozone" : 41, "Solar.R" : 190, "Wind" : 7.4, "Temp" : 67, "Month" : 5, "Day" : 1 }, 
  { "Ozone" : 36, "Solar.R" : 118, "Wind" : 8, "Temp" : 72, "Month" : 5, "Day" : 2 } ]'
\end{Verbatim}

Or equivalently:

\begin{Verbatim}[frame=single]
curl https://cran.ocpu.io/reshape2/R/melt/json -H "Content-Type: application/json" \
  -d '{"id" : ["Month", "Day"], "data" : [ 
    { "Ozone" : 41, "Solar.R" : 190, "Wind" : 7.4, "Temp" : 67, "Month" : 5, "Day" : 1 }, 
    { "Ozone" : 36, "Solar.R" : 118, "Wind" : 8, "Temp" : 72, "Month" : 5, "Day" : 2 } 
  ] }'
\end{Verbatim}

This request basically executes the following R code:

\begin{knitrout}
\definecolor{shadecolor}{rgb}{0.969, 0.969, 0.969}\color{fgcolor}\begin{kframe}
\begin{alltt}
\hlstd{mydata} \hlkwb{<-} \hlstd{airquality[}\hlnum{1}\hlopt{:}\hlnum{2}\hlstd{, ]}
\hlstd{y} \hlkwb{<-} \hlstd{reshape2::}\hlkwd{melt}\hlstd{(}\hlkwc{data} \hlstd{= mydata,} \hlkwc{id} \hlstd{=} \hlkwd{c}\hlstd{(}\hlstr{"Month"}\hlstd{,} \hlstr{"Day"}\hlstd{))}
\hlkwd{cat}\hlstd{(}\hlkwd{toJSON}\hlstd{(y))}
\end{alltt}
\end{kframe}
\end{knitrout}

\end{appendices}
\newpage

\bibliographystyle{plainnat}
\bibliography{json-mapping}
\end{document}